\documentclass[a4paper,11pt]{article}
\pdfoutput=1 

\usepackage{jcappub} 
\usepackage[T1]{fontenc} 

\title{\boldmath Calculation of primordial abundances of light nuclei including a heavy sterile neutrino}

\author[a,b]{M. E. Mosquera,}
\author[b,1]{O. Civitarese\note{Corresponding author.}}

\affiliation[a]{Facultad de Ciencias Astron\'{o}micas y Geof\'{\i}sicas, Universidad Nacional de La Plata, \\ Paseo del Bosque, (1900) La Plata, Argentina}
\affiliation[b]{Department of Physics, University of La Plata, \\ c.c. 67 (1900), La Plata, Argentina}

\emailAdd{mmosquera@fcaglp.unlp.edu.ar}
\emailAdd{osvaldo.civitarese@fisica.unlp.edu.ar}

\abstract{We include the coupling of a heavy sterile neutrino with active neutrinos in the calculation of primordial abundances of light-nuclei. We calculate neutrino distribution functions and primordial abundances, as functions depending on a renormalization of the sterile neutrino distribution function $(a)$, the sterile neutrino mass $(m_s)$ and the mixing angle $(\phi)$. Using the observable data, we set constrains on these parameters, which have the values $a < 0.60$, $\sin^2 \phi=0.15$ and $m_s \approx 4$ keV, for a fixed value of the baryon to photon ratio. When the baryon to photon ratio is allowed to vary, its extracted value is in agreement with the values constrained by Planck observations and by the Wilkinson Microwave Anisotropy Probe (WMAP). It is found that the anomaly in the abundance of $^7$Li persists, in spite of the inclusion of a heavy sterile neutrino.}

\begin{document}
\maketitle
\flushbottom

\section{Introduction}
\label{intro}

The observational data obtained by the WMAP Collaboration \cite{larson11} and by Planck \cite{planck13} signalled the existence of a deficit in the abundance of primordial lithium, which cannot be explained in the context of the standard Big Bang Nucleosynthesis (BBN). Several authors have studied the problem from different points of view: i) the turbulent transport in the radiative zones of stars \cite{richard05}; ii) the existence of a stellar lithium depletion that depends on the mass of the star \cite{melendez10,lind10}; iii) the nuclear physics aspects of the abundance of $^7$Li \cite{kirsebom11,broggini12,civitarese12}; iv) variation of fundamental constants \cite{landau06,landau08,mosquera11b,mosquera10,civitarese10}, among others, but the question is still open since none of these possible explanations, although plausible,  provides a complete solution to the problem.

In previous works, we have analysed the effect of the inclusion of light sterile-neutrinos during the first three minutes of the Universe \cite{civitarese08a,civitarese08b,mosquera11,mosquera14}, to calculate primordial abundances as a function of the active-sterile mixing parameters in the two-state scheme, the 3 + 1 scheme and in the 3 + 2 scheme. We have also analysed the case where the sterile-neutrino might have a variable normalization constant in its occupation factor \cite{mosquera14}. The results of the calculations indicate that the value of the normalization constant should be of the order of (or smaller than)  $0.65$ and that the mixing angle must be zero, in order to be consistent with the observational data.

In this work, we extent our previous study \cite{mosquera14}, by considering a heavy sterile neutrino coupled to the active ones. To perform the calculation we solve the evolution equation of an expanding Universe, including neutrino oscillations in the decay rates. Using the available observational data we set limits on the sterile-neutrino sector.

This work is organized as follows. In Section \ref{formalismo} we present the formalism and in Section \ref{resultados} we present and discuss the results of primordial abundances as functions of the baryon density,  the sterile neutrino occupation factor, and the active-sterile mixing angle. Finally, in Section \ref{conclusion},  the conclusions are drawn.

\section{Formalism}
\label{formalismo}

The matrix which relates neutrino mass-eigenstates and neutrino flavour-eigenstates is the unitary matrix \cite{abazajian12}
\begin{eqnarray}
 U&=&\left(
\begin{array}{cccc}
c_{12} c_{13} \cos \phi & c_{13} s_{12} \cos \phi & s_{13} \cos \phi & \sin \phi\\
\alpha & \delta & s_{23} c_{13}& 0\\
\epsilon & \lambda & c_{23} c_{13} &0\\
- c_{13}c_{12} \sin \phi&-c_{13}s_{12}\sin \phi&-s_{13}\sin \phi&\cos \phi
\end{array}
\right) \, \, \, , \nonumber
\end{eqnarray}
where $i,j=1,2,3$ denote mass eigenstates, $s_{ij}(c_{ij})$ stands for $\sin \theta_{i} (\cos \theta_{ij})$, $\alpha= -s_{12} c_{23}-s_{13}c_{12}s_{23}$, $\delta=c_{23} c_{12} -s_{13}s_{12}s_{23}$, $\epsilon=s_{23} s_{12}-s_{13}c_{12}c_{23}$, $\lambda=-s_{23} c_{12} -s_{13}s_{12}c_{23}$, and $\phi$ is the mixing-angle of the lowest mass-eigenstate with the sterile neutrino. The inclusion of a sterile-neutrino affects the statistical occupation factor of active neutrinos, quantities which are crucial for the determination of primordial abundances. In order to compute these new factors, one must solve the equation \cite{kirilova05}
\begin{eqnarray}
\left(\frac{\partial f}{\partial t}- {\rm H} p \frac{\partial f} {\partial p}\right)&=& -\imath \left[{\mathcal{H}}_0,f\right] \, \, \, ,
\end{eqnarray}
where $f$ is the $4 \times 4$ matrix of the occupation factors, $t$ is the time, ${\rm H}$ is the expansion rate of the Universe (${\rm H}=\mu_P {\rm T}^2$), and ${\mathcal{H}}_0$ is the unperturbed mass term of the neutrino Hamiltonian in the rest frame. We have assumed that at the temperature ${\rm T}_0 = 5 \, {\rm MeV}$ the occupation factors for all neutrinos in the usual flavour representation (namely electron-neutrino, muon-neutrino and tau-neutrino, respectively) are Fermi-Dirac distributions for massless particles with energy ${\rm E}_\nu=p$ ($c=1$ everywhere). For the sterile neutrino, being massive, we assume that its occupation factor is a Fermi-Dirac distribution with energies $\left({\rm E}_s=\sqrt{m_s^2 + p^2}\right)$. The relationship between the sterile neutrino mass and the mass of the mass-eigenstates, $m_i$, can be written as $m_s=\sum_i m_i \left|U_{4i}\right|^2$. This occupation factor is further renormalized by a constant factor $a$ \cite{acero09} which varies between $0$ and $1$. The initial condition for the occupation factors in the mass-eigenstates representation is written 
\begin{eqnarray}
\label{ci} \left. \left(
\begin{array}{cccc}
f_{11}&f_{12}&f_{13}&f_{14}\\
f_{21}&f_{22}&f_{23}&f_{24}\\
f_{31}&f_{32}&f_{33}&f_{34}\\
f_{41}&f_{42}&f_{43}&f_{44}\\
\end{array}
\right)\right|_{T_0}
&=&
\frac{1}{1+e^{p/T_0}} \left(
\begin{array}{cccc}
\cos \phi^2&0&0&-\cos \phi\sin \phi \\
0&1&0&0\\
0&0&1&0\\
-\cos \phi\sin \phi &0&0& \sin \phi^2 \\
\end{array}
\right)\nonumber \\
&&+
\frac{a}{1+e^{E_s/T_0}}\left(
\begin{array}{cccc}
\sin \phi^2&0&0&\cos \phi\sin \phi \\
0&0&0&0\\
0&0&0&0\\
\cos \phi\sin \phi &0&0& \cos \phi^2 \\
\end{array}
\right)\, \, \, .\nonumber \\
\end{eqnarray}
The solutions, for flavour-eigenstates, are 
\begin{eqnarray}
f_{ee}&=& \frac{1}{1+e^\frac{p}{T}} \left\{1+\frac{1}{2}\cos^2\theta_{12} \cos^2\theta_{13}\sin^2 2\phi  \frac{e^{\frac{p}{T}-\beta_0}-1} {1+e^{-\beta_0}} \left[1 -\cos \gamma\right]\right\}\nonumber\\
&&-\frac{1}{2}\cos^2\theta_{12}\cos^2\theta_{13} \sin^2 2\phi \left[1 -\cos \gamma\right] \frac{1-a}{1+e^{\beta_0}} \, \, \, ,\nonumber\\
f_{\mu\mu}&=& \frac{1}{1+e^\frac{p}{T}} \left\{1+\frac{1}{2}\sin^2\theta_{12} \cos^2\theta_{13} \sin^2 2\phi  \frac{e^{\frac{p}{T}-\beta_0}-1}{1+e^{-\beta_0}} \left[1 -\cos \gamma \right]\right\}\nonumber\\
&&-\frac{1}{2}\sin^2\theta_{12}\cos^2\theta_{13} \sin^2 2\phi \left[1 -\cos \gamma\right] \frac{1-a}{1+e^{\beta_0}} \, \, \, ,\nonumber\\
f_{\tau\tau}&=& \frac{1}{1+e^\frac{p}{T}}\left\{1+\frac{1}{2} \sin^2\theta_{13}\sin^2 2\phi  \frac{e^{\frac{p}{T}-\beta_0}-1}{1+e^{-\beta_0}} \left[1 -\cos \gamma \right]\right\}\nonumber\\ 
&&-\frac{1}{2}\sin^2\theta_{13}\sin^2 2\phi \left[1 -\cos \gamma\right] \frac{1-a}{1+e^{\beta_0}}\, \, \, , \nonumber\\
f_{ss}&=& \frac{1}{1+e^{\beta_0}}\left\{-\frac{1}{2}\sin^2 2\phi \frac{e^{\beta_0-\frac{p}{T}}-1}{1+e^{-\frac{p}{T}}} \left[1 -\cos \gamma\right]\right\}\nonumber\\ 
&&+\frac{1}{2}\sin^2 2\phi \frac{\lambda}{1+e^{\beta_0}} \left[1 -\cos \gamma\right] +\frac{a}{1+e^{\beta_0}} \, \, \, ,
\end{eqnarray}
where
\begin{eqnarray}
\gamma &=&\frac{1}{\mu_P}\left[-\frac{m_1^2 T}{6 p}\left(\frac{1}{T^3}-\frac{1}{T_0^3} \right) + \frac{p}{T} \left(\frac{1}{T}-\frac{1}{T_0}\right)- \frac{1}{2 T^2}\sqrt{m_4^2+p^2 } +\frac{1}{2 T_0^2} \sqrt{m_4^2+\left(\frac{p T_0}{T}\right)^2}\right.\nonumber \\
&& \left. \hskip 0.8cm-\frac{p^2}{2 m_4 T^2} \ln \left(\frac{T_0}{T} \frac{m_4 + \sqrt{m_4^2+p^2}}{m_4+\sqrt{m_4^2+ \left(\frac{p T_0}{T}\right)^2}} \right)\right], \nonumber
\end{eqnarray}
and $\beta_0 = T_0^{-1}\sqrt{m_s^2+\left(\frac{p T_0}{T}\right)^2}$.

\section{Results}
\label{resultados}

In order to obtain the primordial abundances as functions of the active-sterile neutrino mixing parameters, we have modified the numerical code of Kawano \cite{kawano88,kawano92}. The active neutrino mixing parameters were extracted  from the SNO, SK, GNO, CHOOZ, DAYA BAY and DOUBLE CHOOZ experiments \cite{sno02,k2k06,gno05,chooz09,dayabay,doublechooz}. The light-neutrino mass was fixed at the square root of the lowest squared mass difference. The active-sterile mixing angle was fixed at the value $\sin^2 2\phi=0.15$. We have considered two different cases; (i) by fixing the baryon density at the value determined from WMAP \cite{larson11}, and (ii) varying $\eta_B$.

To obtain the best value for the parameters of the sterile-neutrino sector we have performed a $\chi^2$ minimization in the corresponding parametric space. The observational data for deuterium have been extracted from Refs. \cite{balashev10,pettini08,cooke14,noterdaeme12,pettini12}. We use the data from Refs.  \cite{izotov10,aver12,peimbert07,villanova09,aver13,izotov13} for $^4$He and, for $^7$Li we have considered the data given by Refs. \cite{nissen12,sbordone10,lind09,monaco12}. Regarding the consistency of the data, we have followed the treatment of Ref. \cite{pdgbook} and increased the errors by a fixed factor $\Theta_{^4{\rm He}}=1.30$, for the other cases the errors were not changed.

\subsection{Results with $\eta_B$ fixed}

\begin{figure}[t]
\centering
\includegraphics[width=.80 \textwidth,origin=c,angle=0]{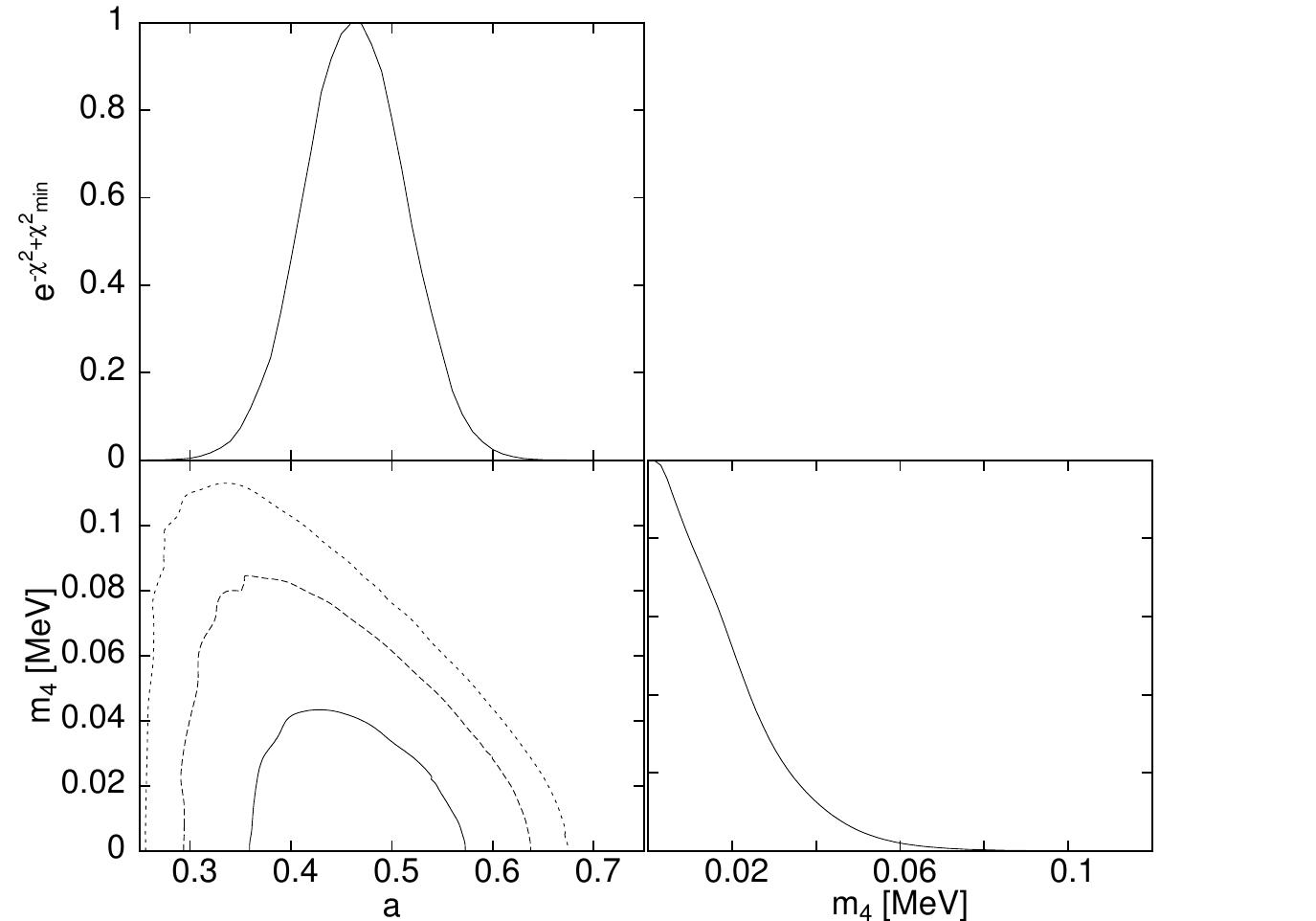} 
\hfill
\caption{\label{todo} Likelihood and 1$\sigma$, 2$\sigma$ and 3$\sigma$ contour plots for the parameter $a$ (renormalization of the sterile neutrino occupation factor) and the sterile-neutrino mass, when all data are considered in the $\chi^2$-test. The mixing angle $\phi$ has been fixed at the value $\sin^2 2\phi=0.15$.}
\end{figure}

\begin{figure}[t]
\centering
\includegraphics[width=.80\textwidth,origin=c,angle=0]{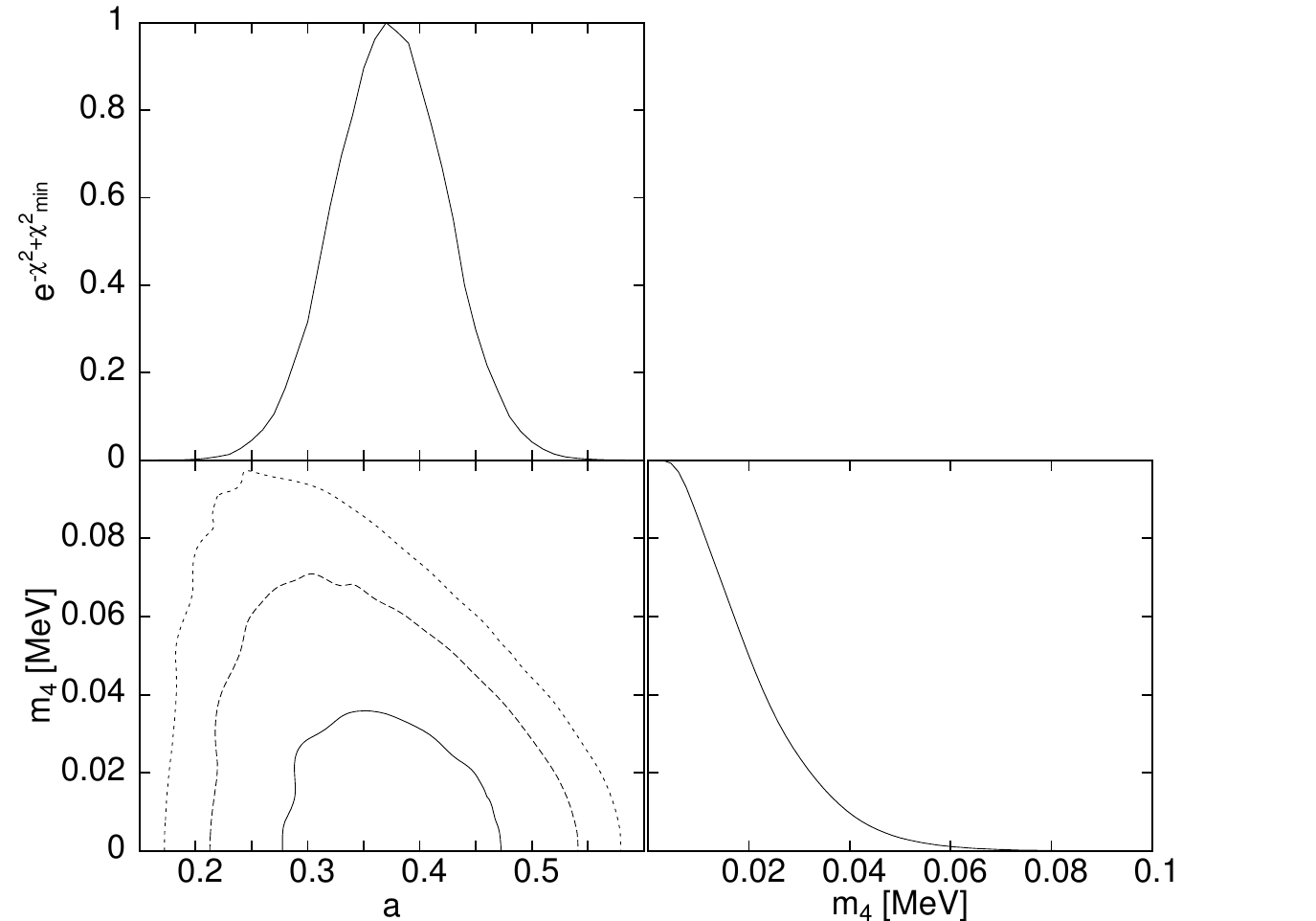} 
\caption{\label{sinli} Likelihood and 1$\sigma$, 2$\sigma$ and 3$\sigma$ contour plots, in the same notation of Figure 1, when the data of $^7$Li are not considered in the $\chi^2$-test. The mixing angle $\phi$ has been fixed at the value $\sin^2 2\phi=0.15$.}
\end{figure}

In this section we present the results of the calculation of primordial abundances performed as a function of two parameters: the sterile-neutrino mass and the renormalization factor $a$. The value of baryon density was fixed at the WMAP value. We performed a $\chi^2$-analysis in order to obtain the best-fit value of the parameters. The results are the following
\begin{itemize}
\item All data
\begin{eqnarray}
a&=& 0.46^{+0.08}_{-0.06}\, \, \, ,\nonumber \\
m_4 &=& 0.003^{+0.022}_{-0.003} \,\, {\rm MeV}\, \, \, ,\nonumber \\
\chi^2/(N-2)&=& 10.76\, \, \, .
\end{eqnarray}
\item All data but $^7$Li
\begin{eqnarray}
a &=&0.37 \pm 0.06\, \, \, ,\nonumber \\
m_4 &=&0.004^{+0.021}_{-0.004} \,\, {\rm MeV}\, \, \, , \nonumber \\
\chi^2/(N-2)&=& 3.23\, \, \, .
\end{eqnarray}
\end{itemize}
The first set of results have been obtained by taking all data on primordial abundances, and the second one is the set of results obtained by removing the data on the abundance of lithium. Figures \ref{todo} and \ref{sinli} show the contour plots in the parametric-plane $(m_4,a)$ and the likelihood contour plots. 

The statistical analysis shows that the data on lithium  may not be consistent with the other data on primordial abundances, since $\frac{\chi^2(\rm{with \;Li})}{\chi^2(\rm{without \; Li})} \approx 3$. However, both sets yield almost the same value for the sterile-neutrino mass and comparable values for the renormalization factor of the sterile-neutrino occupation.

\subsection{Results with $\eta_B$ variable}

The results obtained by allowing the variation of the baryon to photon ratio, $\eta_B$, are the following:
\begin{eqnarray}
\eta_B &=&(6.20_{-0.07}^{+0.06})\times 10^{-10}\, \, \, , \nonumber \\
a &=& 0.53 \pm 0.09\, \, \, ,\nonumber \\
m_4 &=& 0.000^{+0.040} \, \,{\rm MeV}\, \, \, ,\nonumber \\
\chi^2/(N-3)&=& 11.40\, \, \, .
\end{eqnarray}

\begin{figure}[t]
\centering
\includegraphics[width=1.0\textwidth,origin=c,angle=0]{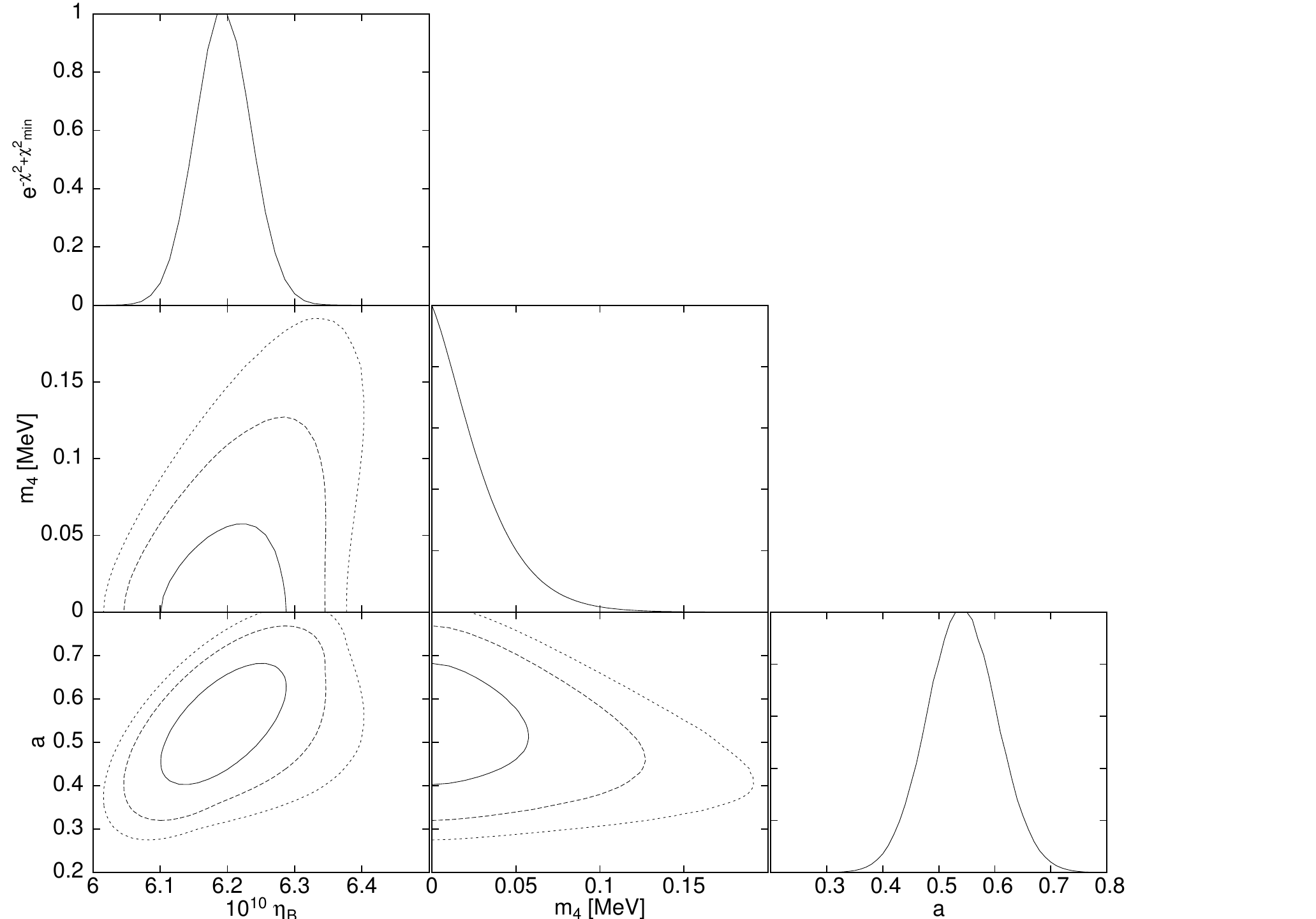} 
\caption{\label{todo-eta-1} Likelihood and 1$\sigma$, 2$\sigma$ and 3$\sigma$ contour plots for $\eta_B$, $m_4$ and the parameter $a$. The mixing angle $\phi$ has been fixed at the value $\sin^2 2\phi=0.15$.}
\end{figure}

Figure \ref{todo-eta-1} shows the contour plots resulting from the calculations. The extracted value for $\eta_B$ is in good agreement with data, the renormalization parameter $a$ is somehow larger than the one obtained with a fixed value of $\eta_B$ \cite{larson11,planck13}, but the fit favors a massless sterile-neutrino. The $\chi^2$-value for this case is similar to the one corresponding to the minimization with a fix $\eta_B$ when all primordial species are considered.  Since we have three parameters to adjust and three set of data, we cannot, for the case of variable baryon to photon ratio, excluded the data on lithium. 

\section{Conclusion}
\label{conclusion}

The presence of an extra heavy-neutrino affects the primordial abundances produced during the first three minutes of the Universe. We have calculated the occupation factors for active and sterile-neutrinos (3+1 scheme) and the neutron to proton decay rates as a function of the new mass eigenstate, the active sterile neutrino mixing angle, and the parameter $a$, in order to obtain the primordial abundances of deuterium, helium and lithium. As in previous works, we have found a sensitivity of the abundances to the active sterile neutrino mixing \cite{kishimoto06,civitarese08a,civitarese08b,mosquera11}. The value for the parameter $a$ remains lower than $0.60$ at 1$\sigma$, in agreement with previous calculations \cite{mosquera14}. The results suggest that the primordial abundances are more affected by changes in $a$ and in the neutrino mass rather than in the mixing angle between active and sterile neutrino species.

\acknowledgments

Support for this work was provided by the National Research Council (CONICET) of Argentina, and by the ANPCYT of Argentina. The authors are members of the Scientific Research Career of the CONICET.

\end{document}